# Synthesis and water permeation studies of polysulfone based composite membranes having vertically aligned CNTs

Bhakti Hirani[1,2] and P.S. Goyal[1]
[1]Department of Electronics, Pillai College of Engineering,
New Panvel - 410206, India
[2]Department of Biotechnology, Pillai College of Arts, Commerce and Science (Autonomous)
New Panvel - 410206, India

## Abstract

Polymeric membranes, including Polysulfone (PSf) membranes, are routinely used for water treatment. It is known for quite some time that water permeability of above membranes can be improved if one incorporates carbon nanotubes (single-walled, SWCNTs or multi-walled, MWCNTs) in to the membrane and aligns them in direction of flow of water. This paper reports a method of synthesizing polymeric membranes having vertically aligned hollow CNTs embedded in them. This involves mixing of nanomagnetic particles in the dope solution and casting of membrane in presence of moderate magnetic fields.

A semi-automatic membrane casting machine which allows casting of membrane in presence magnetic field was designed and fabricated. PSf nanocomposite membranes, having vertically aligned MWCNTSs, were synthesized using above machine. The effect of magnetic field and the exposure time on the water permeation of above membranes was studied. It was seen that water permeability of membrane increases by a factor of 4 when the magnetic field is increased from 0 to 1500 Gauss. There was additional 40% increase in water permeability, when the time for which film was exposed to magnetic field was increased from 5 sec. to 10 sec.

## 1. Introduction

Membranes are porous polymeric films, and are used in many fields such as waste water treatment, industrial separation processes and protein recovery etc[1-3]. Depending on the pore size, water treatment processes employ several types of membranes. Ultra-filtration (UF) membranes have pore sizes in range of about 5-100 nm and they are used for purification of brackish water. These membranes can reject large particles, micro-organisms, bacteria and soluble macromolecules such as proteins. Efforts are being made world over, to improve the water permeability of above membranes. One of the methods for enhancing the water permeability of above membranes involves incorporating carbon nano-tubes (CNTs) into them and aligning CNTs in a direction perpendicular to the surface of the membrane[4-13]. Fig.1 is a schematic drawing of CNT based UF membrane. So far, no satisfactory method of aligning CNTs has been developed which could be exploited for commercial production of above membranes. This paper reports a method of synthesizing polysulfone composite UF membranes having vertically aligned multi-walled carbon nanotubes (MWCNTs).

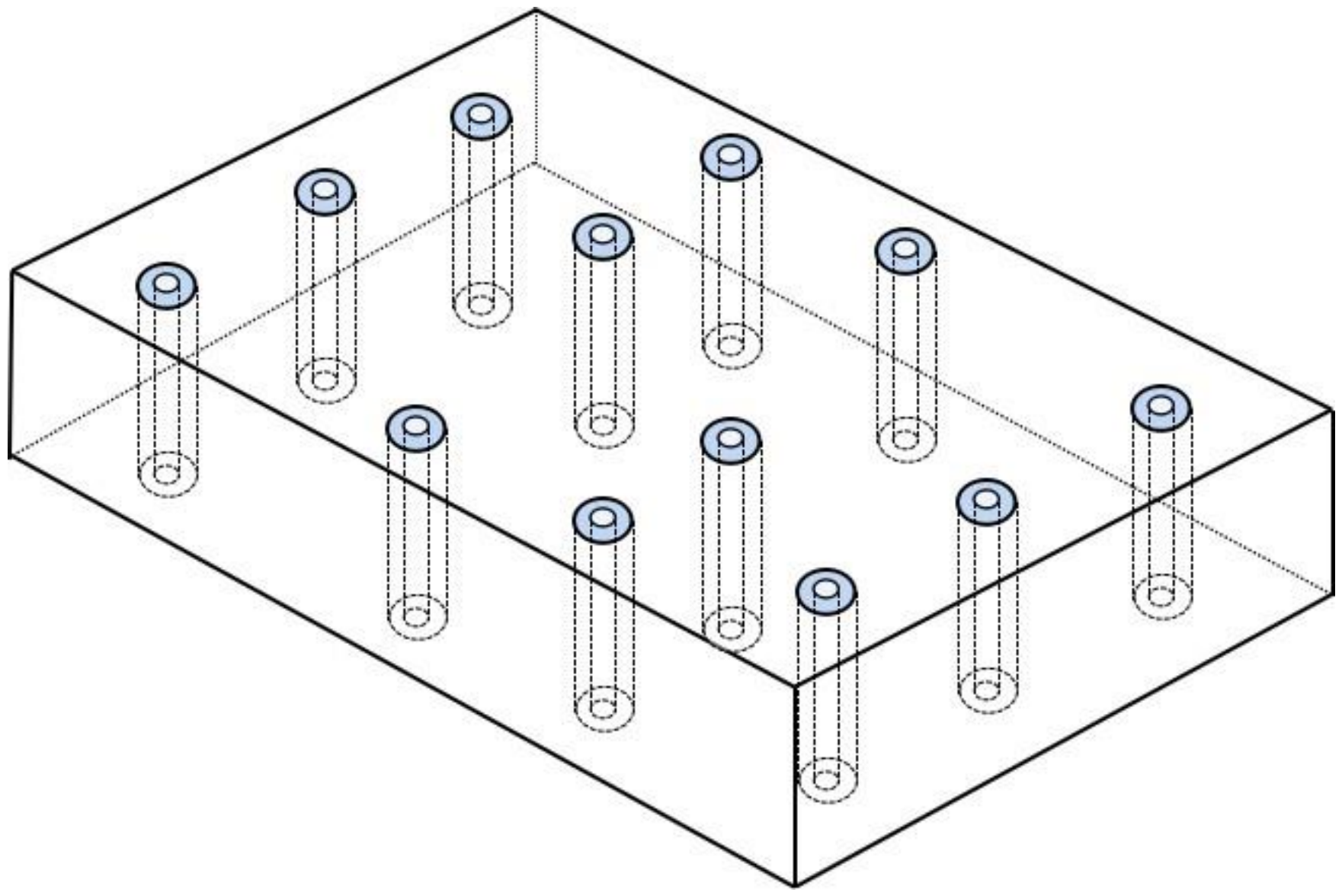

**Fig. 1. Artistic view of the carbon nanotubes-based nanocomposite membrane**

Polysulfone (PSf) is a commonly used membrane material[14-17]. In addition to PSf, organic polymer polyvinyl pyrrolidone (PVP) is added to the to the cast solution for improving hydrophilicity, porosity and antifouling nature of resultant membrane. Carbon nanotubes (CNTs) are cylinders of one or more layers of graphene[18]. They are referred to as single-wall Nanotubes (SWCNTs) or multiwall nanotubes (MWCNT) depending on whether cylinder consists of single or multi-layers of grapheme. Diameters of SWCNTs and MWCNTs are typically 0.8 to 2 nm and 5 to 20 nm respectively. The lengths of CNTs could range from 100 nm to several mm. CNTs are widely used for various devices and as additives or reinforcing materials in advanced composites[18-22]. In most of these applications, it is desirable that CNTs are aligned along a suitable direction[19-21]. A variety of methods for aligning CNTs have been developed, and they have their merits or demerits depending on the application[23 - 29].

There are several studies which deal with synthesis of CNT based nanocomposite membranes[4-13]. Majumder *et al.*[5], for example, synthesized membranes having aligned SWCNTs and showed that alignment of SWCNTs leads to higher water permeability. Water molecules move through inner-tubular cavity of nanotubes and provide additional contribution to the permeability of the membrane. Srivastava *et al.*[6] synthesized hollow carbon cylinder comprising radially aligned SWCNTs and demonstrated that the system can be used for removing salt from water. The above studies exploit the fact that CNTs get aligned during synthesis, if the CNTs are synthesized using process of chemical vapor deposition (CVD). This method of aligning CNTs, however, cannot be upgraded to commercial production of membranes. The other method of aligning CNTs involves use of magnetic field. It is seen, however, that one has to use very large magnetic flux densities ((B

> 10T) for achieving reasonable alignment of CNTs. The method suggested by Kordas *et.al.*[28] uses small magnetic fields, but can't be used for present application as it involves entrapping of magnetic nanoparticles in the inner-tubular cavity of nanotubes. Goyal *et al.*[30] suggested that it should be possible to align CNTs (SWCNTs or MWCNTs) and improve water permeability of membranes if, in addition to CNTs, the dope solution contains magnetic nanoparticles (e.g $Fe_3O_4$ nanoparticles) and the membranes are casted in presence of magnetic field. This method of synthesizing CNT based membrane has the potential of up scaling. This paper reports results of water permeability studies on Polysulfone - MWCNT composite membranes where MWCNTs were aligned using above method.

The next section deals with the materials and chemicals used in synthesis of membranes. This section also gives details of synthesis and characterization of nanomagnetic particles of $Fe_3O_4$. Section 3 deals with Polysulfone-MWCNT (un-aligned) membranes, where no effort was made to align the CNTs. In addition to synthesis details, this section gives results of water permeation studies for the above membranes. The details of cross-filtration cell and the method of measuring water permeability are discussed. Section 4 deals with Polysulfone-MWCNT (aligned) membranes. The details of the method of aligning MWCNTs are given in Section 4.1. Section 4.2 deals with Semi-automatic Membrane Casting Machine which has been specially developed for the purpose; it allows casting of membrane in presence of magnetic field. Section 4.3 deals with synthesis and water permeability studies on Polysulfone-MWCNT (aligned) membranes. Section 5 gives a summary.

## 2. Materials

### *2.1 Chemicals*

Polysulfone (PSf, Udel P 3500 LCD MB 7, MW = 40,000 Dalton), polyvinyl pyrrolidone (PVP, K-30; MW: 40,000 Dalton) and N-methyl – 2 – pyrrolidone (NMP, AR Grade, minimum assay of 99.5%) were used in the cast solution as the base polymer, the pore former and the solvent respectively[14-17]. MWCNTs and nanoparticles of $Fe_3O_4$ were used as additives. PSf was obtained from M/s. Solvay Specialties India Pvt. Ltd. NMP and PVP were purchased from Sisco Research Laboratory, Pvt. Ltd. India and MWCNTs (OD: 6 - 13 nm, Length: 2.5 - 20μm) were purchased from M/s. United Nanotech Innovations Pvt. Ltd. India. The above chemicals have been used as obtained. Oleic acid coated $Fe_3O_4$ nanoparticles were synthesized in our laboratory. The chemicals for this (ferrous chloride 98%, ferric chloride reagent grade, 97%) and ammonium hydroxide solution (28.0 – 30.0%) were purchased from Sigma Aldrich. Extra pure oleic acid (MW: 282.47) was purchased from LOBA Chemicals.

### *2.2. Synthesis of Oleic acid coated $Fe_3O_4$ nanoparticles*

The $Fe_3O_4$ nanoparticles were synthesized by chemical co-precipitation method following the protocol, given by Mehta and Upadhyay[31]. 100ml of $FeCl_3$ (0.2M, pH: 3.62) and 100 ml of $FeCl_2$ (0.1M, pH 1.49) were mixed and stirred gently for 10 min using mechanical stirrer. 400ml of Ammonia solution (0.8M) was slowly dropped into the solution with continuous mechanical stirring for 15 mins until the pH reaches 9.8. The following chemical reaction results in precipitation of $Fe_3O_4$ and these precipitates exist as nanoparticles under above experimental conditions.

$$2FeCl_3 + FeCl_2 + 8NH_3 + 4H_2O \rightarrow Fe_3O_4 + 8NH_4Cl$$

The $Fe_3O_4$ nanoparticles were separated using a strong magnet and were washed thrice with 100 ml of hot distilled water. 100 ml of 0.8M Ammonia solution was added to above magnetic slurry and stirred continuously with mechanical stirrer at 50°C. Then, 2.25 ml of oleic acid was added dropwise to the above solution. The suspension was stirred and dispersed at 50°C for 1 hr. and temperature as allowed to rise up to 90°C. The solution was maintained at 90°C for 5 min and then allowed to cool. When solution attains room temperature, dilute HCl was added to obtain the pH of ~4.5. The magnetic slurry so obtained was washed with 150 ml of acetone to remove the impurities. Finally, the magnetic slurry was allowed to dry at room temperature to obtain the powder. This way, around 2 grams of $Fe_3O_4$ nanoparticles were obtained. It may be mentioned that oleic acid molecules get attached to nanoparticle and prevent aggregation of $Fe_3O_4$ particles. The salient features of our work on synthesis and characterization of $Fe_3O_4$ nanoparticles were reported at a conference[32].

*2.3. Characterization of Oleic acid coated $Fe_3O_4$ nanoparticle*

X-ray diffraction pattern (Fig.2) for a powder of $Fe_3O_4$ nanoparticles was recorded at UGC DAE CSR, Indore using Bruker D8 diffractometer with Cu Kα radiation source (λ = 0.154 nm). There are no extra peaks and data show that nanoparticles are made of single phase $Fe_3O_4$. The peaks at (220), (311), (400), (422), (333) and (440) in the measured pattern are consistent with the standard pattern (JCPDS Card No. 74-0748). The lattice parameter of a = 8.3310 Å, obtained from present data is very close to the published value[33]. Further, it is seen Bragg peaks are broad and this is consistent with the fact that $Fe_3O_4$ exists in form of nanoparticles. The sizes of nanoparticles were obtained from width of Bragg peaks using the Debye-Scherrer formula. The average size of the $Fe_3O_4$ nanoparticles as obtained from width of (311) peak is about 12 nm.

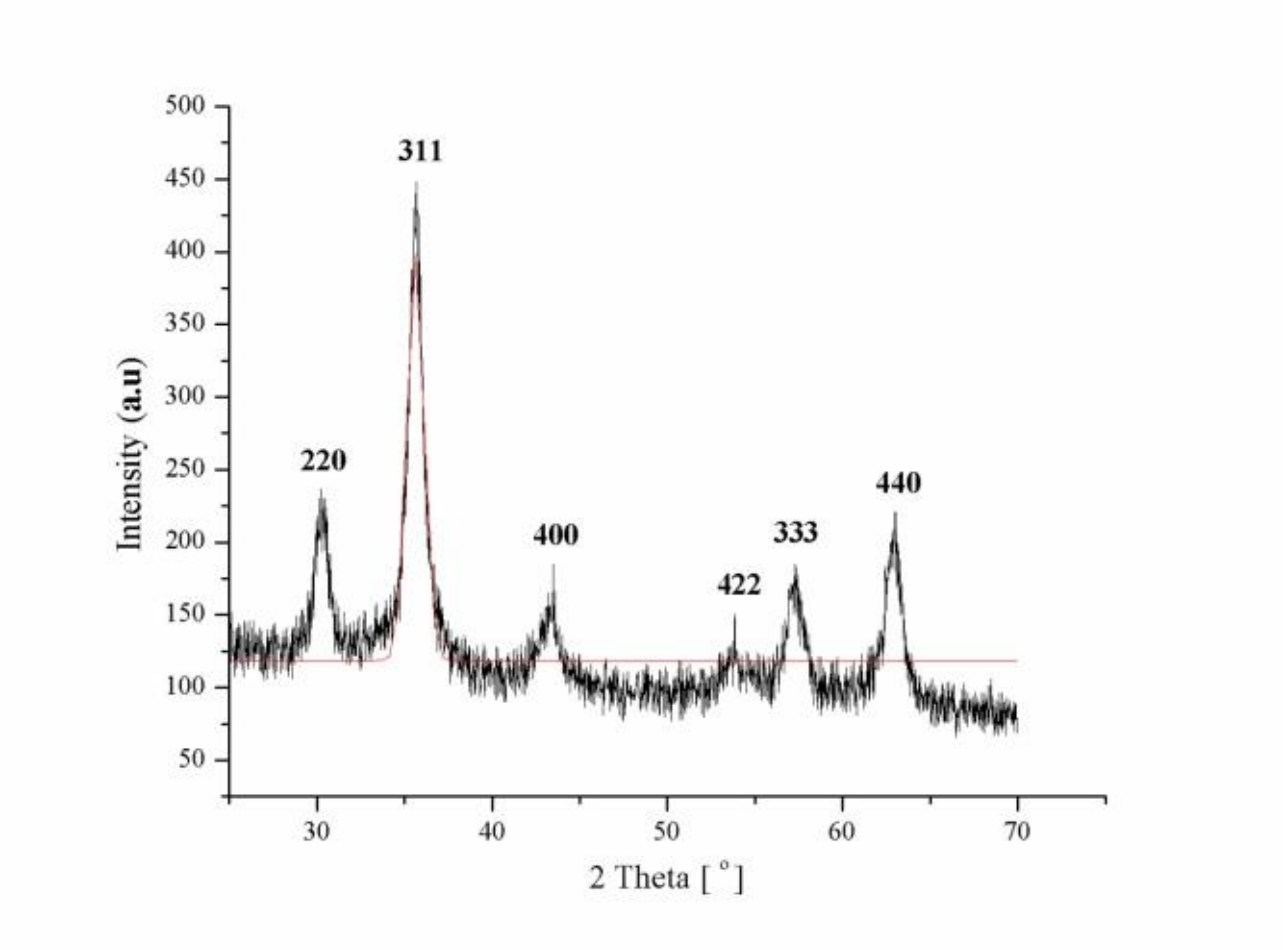

**Fig.2. XRD pattern of oleic acid coated $Fe_3O_4$ nanoparticles**

The sizes of nanoparticles were independently determined using PHILIPS Transmission electron microscope (operating at voltage 200kv) at SAIF, IIT Mumbai. Fig.3 is a TEM picture of above $Fe_3O_4$ nanoparticles. It is seen that particles are nearly spherical and there is a distribution in sizes. The inset in Fig.3 shows the size distribution of nanoparticles. It is noted that sizes of most of the particles are in range of 10-14 nm. This is in agreement with XRD data which gave an average size of 12 nm.

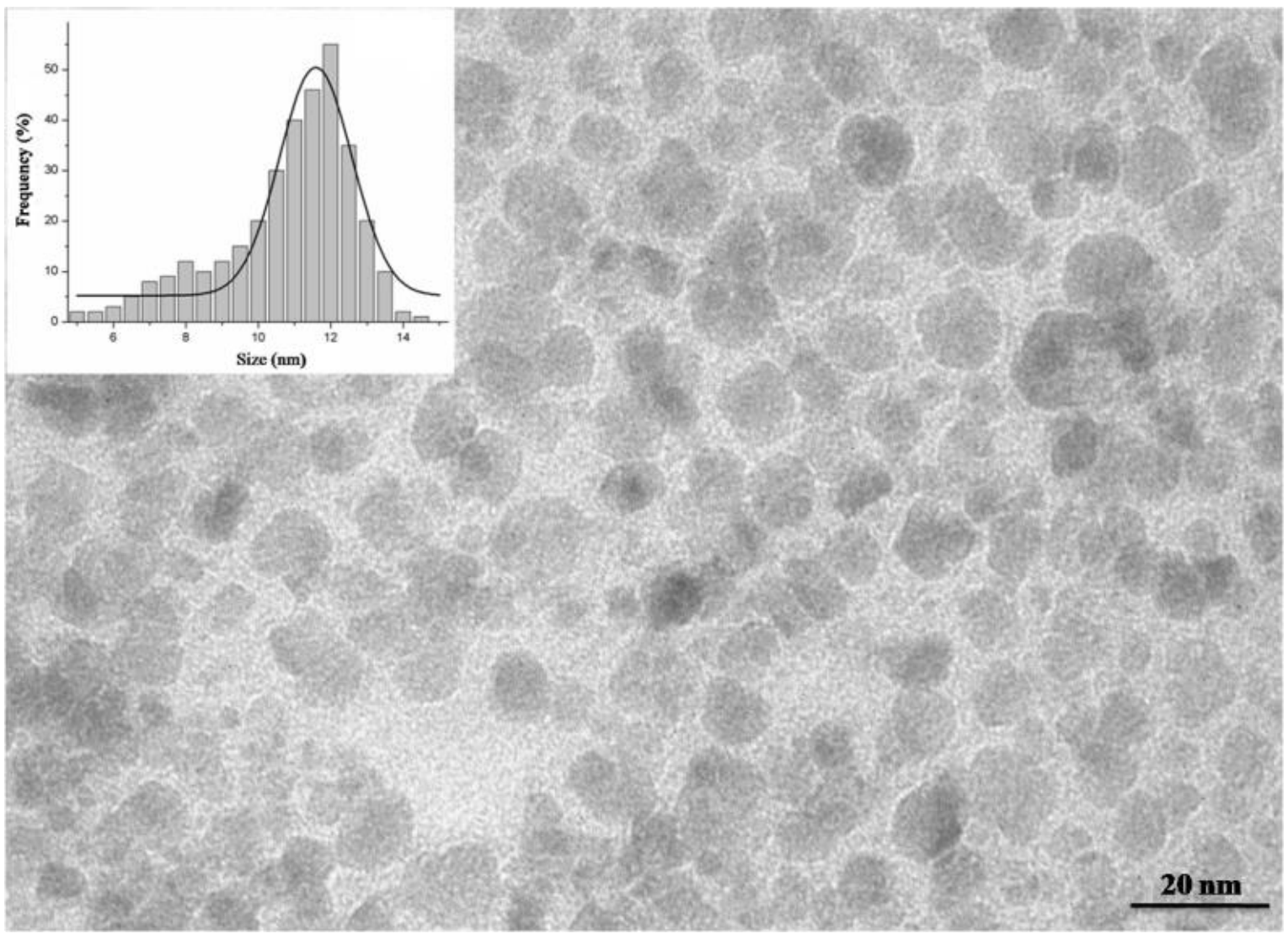

**Fig.3. TEM of $Fe_3O_4$ nanoparticles. The inset shows the size distribution of particles.**

In addition to above, comparative FTIR studies were carried out on oleic acid coated nanoparticles and pure oleic acid. The results were similar to those reported in earlier paper[32]. In particular, it was seen that 1703 $cm^{-1}$ peak (arising from axial C=O stretching) for pure oleic acid shifts to 1526 $cm^{-1}$ for nanoparticle system. This indicates that oleic acid molecules are attached to nanoparticles[34].

## 3. Polysulfone-MWCNT (un-aligned) nanocomposite membranes

*3.1 Synthesis of polysulfone based membranes with and without additives*

The PSf membranes, with and without additives, were synthesized using phase inversion method[35-36]. In all, four types of membranes were synthesized. They consist of (i) Pure PSf, (ii) PSf + MWCNT, (iii) PSf + $Fe_3O_4$ and (iv) PSf + MWCNT + $Fe_3O_4$. The dope solutions were prepared as per the compositions given in Table-1. Nanoadditives (MWCNTs or $Fe_3O_4$) were first dispersed in the solvent (NMP) by sonicating it for 30 minutes. Next, the pore former (PVP) is dissolved and finally the PSf beads are added and the solution was stirred continuously for 3 – 4 days unless a homogenous casting solution was obtained.

Membranes were casted by pouring the dope solution onto the glass plate (~20 cm x 16 cm) and using a casting blade to manually spread the solution to obtain a thin layer of uniform thickness. Glass plate was taped on parallel ends to ensure the resulting membrane layer has thickness of ~200 μm. After casting, the glass plate was immersed into a precipitation tank having distilled water to enable the phase inversion process to happen. The resulting membrane is soaked in distilled water for several days before use. Results of preliminary studies on above membranes were presented at a conference[37,38]. The dope solution containing MWCNTs and $Fe_3O_4$ nanoparticles was used for aligning MWCNTs. (see next section).

*3.2 Water permeation studies on membranes*

The water permeation of above four types of membranes was measured using a cross-flow filtration cell, which was designed and fabricated indigenously. Fig.4 is a schematic drawing and Figures 5a and 5b are photographs of the cross-flow cell. The cross-flow cell allows measurement of the flux of the fluid permeating through membrane, under varying trans-membrane pressures in range of 1 to 10 bars. The experiment involves installing the test membrane in its housing (Fig.5b), allowing the water to pass through the membrane and collecting the permeated water in a beaker for a fixed time. The present set-up allows testing of 35 mm diameter membrane and there is provision to test four membranes simultaneously. The flux F ($LM^{-2}H^{-1}$) of water permeating through the membrane is obtained using the following formula:

$$F = V / St \qquad [1]$$

where V (Liters) is the volume of permeate collected in time t (Hours). S is the active area of the membrane (meter$^2$).

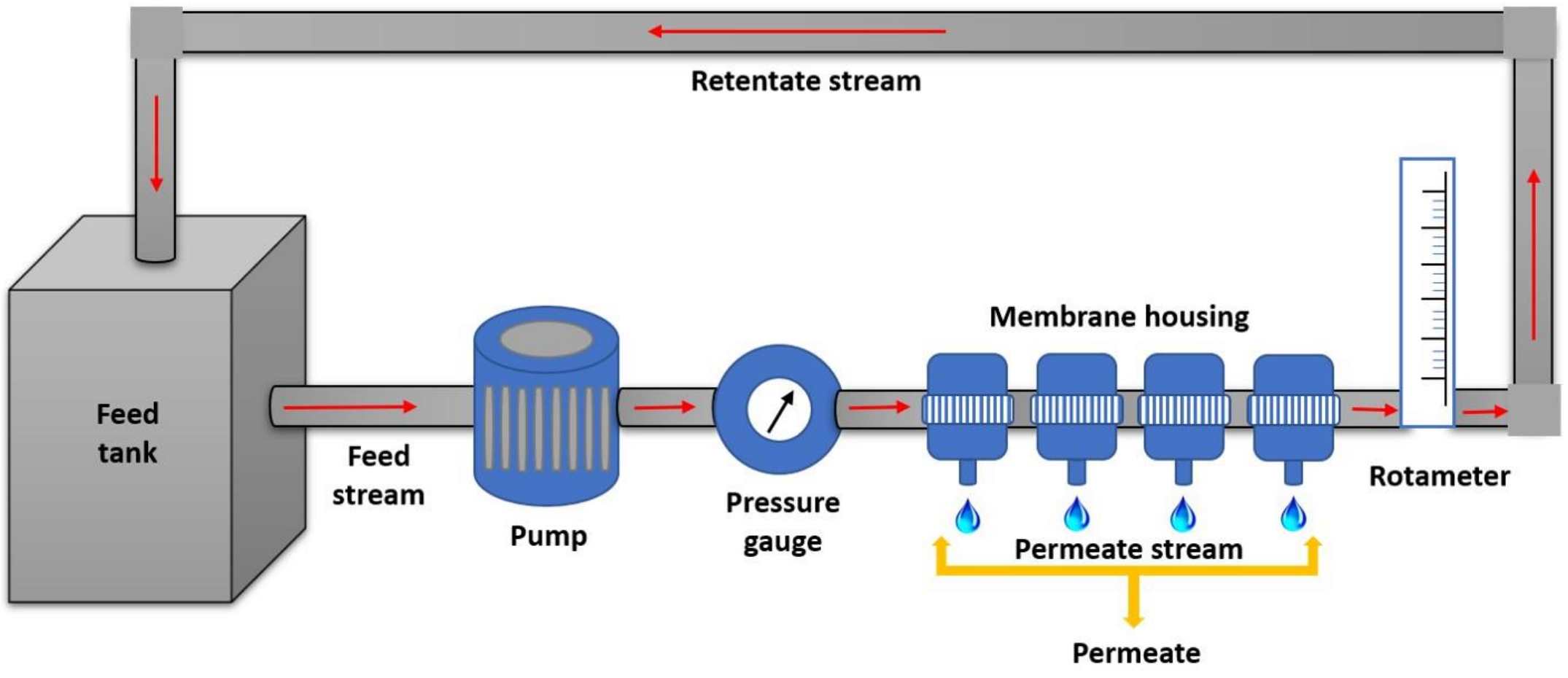

**Fig 4. Schematic diagram of cross-flow filtration cell**

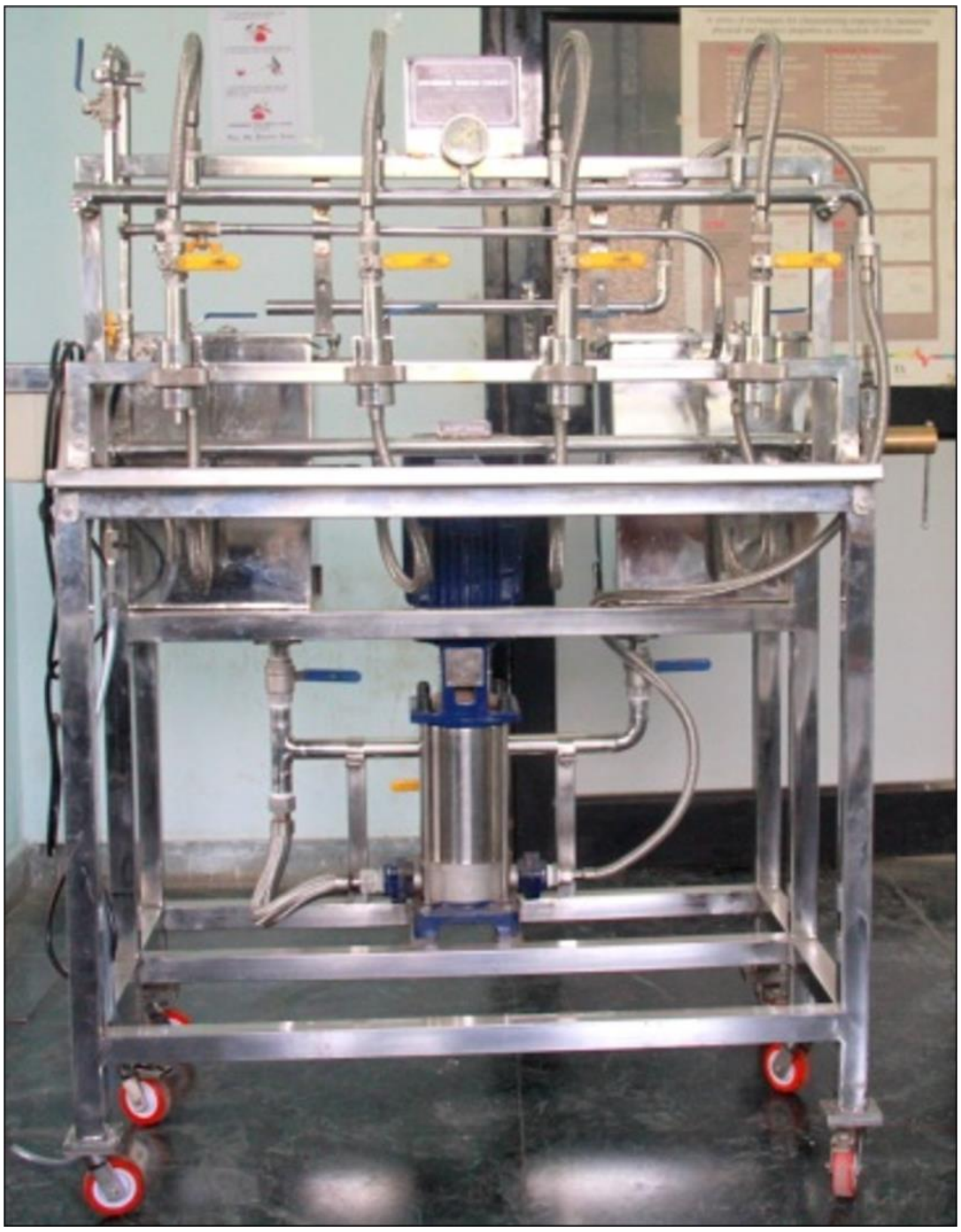

**Fig.5.a Photograph of cross-flow filtration cell**

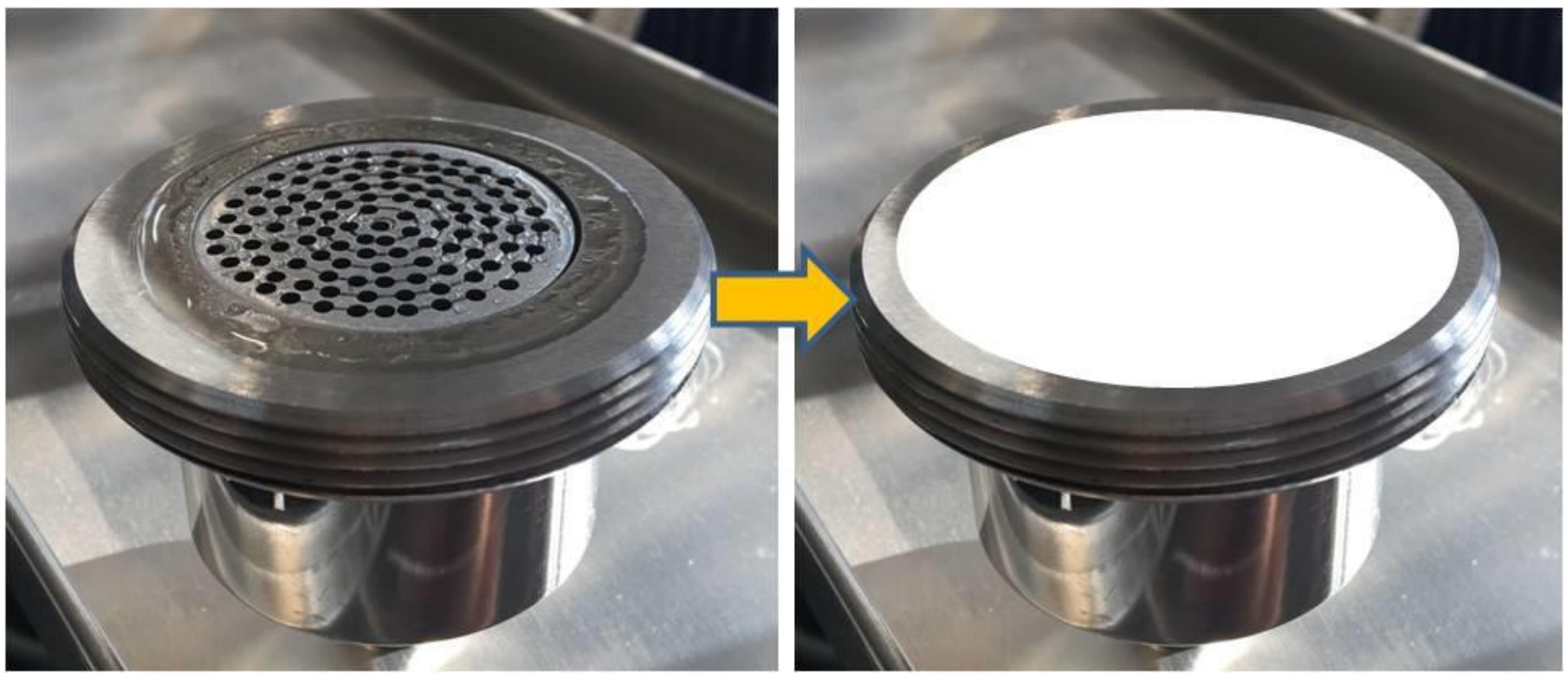

**Fig.5. b. Photograph of the membrane housing**

The permeated water flux was measured for all membranes corresponding to trans-membrane pressure of 2 bar. The membrane was supported in the cell by a porous stainless-steel disk as shown in Fig.5b. The active area of the membrane was 10.17 $cm^2$. The water was allowed to pass through the membrane and permeated water was collected for 5 minutes. The average of four readings were used to obtain the flow rate. The results for all four types of polysulfone composite (un-aligned) membranes are tabulated in Table 1.

**Table 1: Composition, water permeation and rejection of Polysulfone composite (un-aligned) membranes**

| Sr. No. | Sample | Composition of membrane | | | | Flux of permeated water ($LM^{-2}H^{-1}$) | PEO Rejection (%) |
|---|---|---|---|---|---|---|---|
| | | PSf (gm) | PVP (gm) | NMP (ml) | Additives | | |
| 1. | Pure PSf | 25.0 | 7.5 | 87.5 | Nil | 96* | 84 |
| 2. | PSf + MWCNT | 25.0 | 7.5 | 87.5 | MWCNT (0.1 gm) | 288 | 86 |
| 3 | PSf + $Fe_3O_4$ | 25.0 | 7.5 | 87.5 | $Fe_3O_4$ (1.0 gm) | 151 | 85 |
| 4. | PSf + MWCNT + $Fe_3O_4$ | 25.0 | 7.5 | 87.5 | MWCNT (0.1 gm), $Fe_3O_4$ (1.0 gm) | 306 | 83 |

The measured value of permeated water flux F = 96 L $M^{-2}$ $H^{-1}$ for control PSf membrane is reasonable. The values of F obtained by Ravishankar *et al.*[15] and Bedar *et al*[17] for PSf membranes are 52 $LM^{-2}H^{-1}$ and 156 $LM^{-2}H^{-1}$ respectively. The differences in the above values could be because of differences in composition of dope solutions or differences in thicknesses of the membranes. The present studies show that permeated water flux of membrane does not change on addition of $Fe_3O_4$ nanoparticles. The value of F for (PSf + $Fe_3O_4$) membrane is nearly same as that of pure PSf membrane. The permeated water flux for (PSf + MWCNT) membrane is, however, about three times larger as compared to that for pure PSf membrane. The reasons for this increase in permeability of (PSf + MWCNT) compared that for PSf membrane are not clear. Again, it is seen that permeability of (PSf + MWCNT) membrane does not change on addition of $Fe_3O_4$ nanoparticles.

The solute rejection behavior of above membranes was studied using a procedure similar to that used by Bedar *et al.*[17]. It involved dissolving about 2 gm of Polyethylene Oxide (PEO) in 10 liters of ultrapure water and measuring the PEO concentration in the feed and in the permeate. The average molecular weight of PEO was 100 KDa. The concentration of PEO in the feed and in permeate was measured using Total Organic Carbon (TOC) analyzer. The above rejection studies were carried out at Membrane Development Section, BARC, Mumbai. The results are given in Table 1. It is seen that rejection efficiency for PEO is nearly same (~84%) for all the membranes. This value for rejection efficiency is reasonable as it is comparable to the value (94%) obtained by Bedar *et al.* for pure PSf membrane.

*3.3 Study of pore sizes in membranes using Scanning Electron Microscope Small Angle Neutron Scattering*

Scanning electron microscope (SEM) and Small Angle Neutron Scattering (SANS) studies were carried on above membranes to get a better insight into results of permeation and rejection studies. SEM micrograph of pure PSf membrane, as recorded using SEM machine at IIT, Bombay is shown Fig.6. It is seen that membrane has pores which vary from ~10 nm to ~50 nm in their sizes.  The maximum numbers of pores are of about 20 nm.

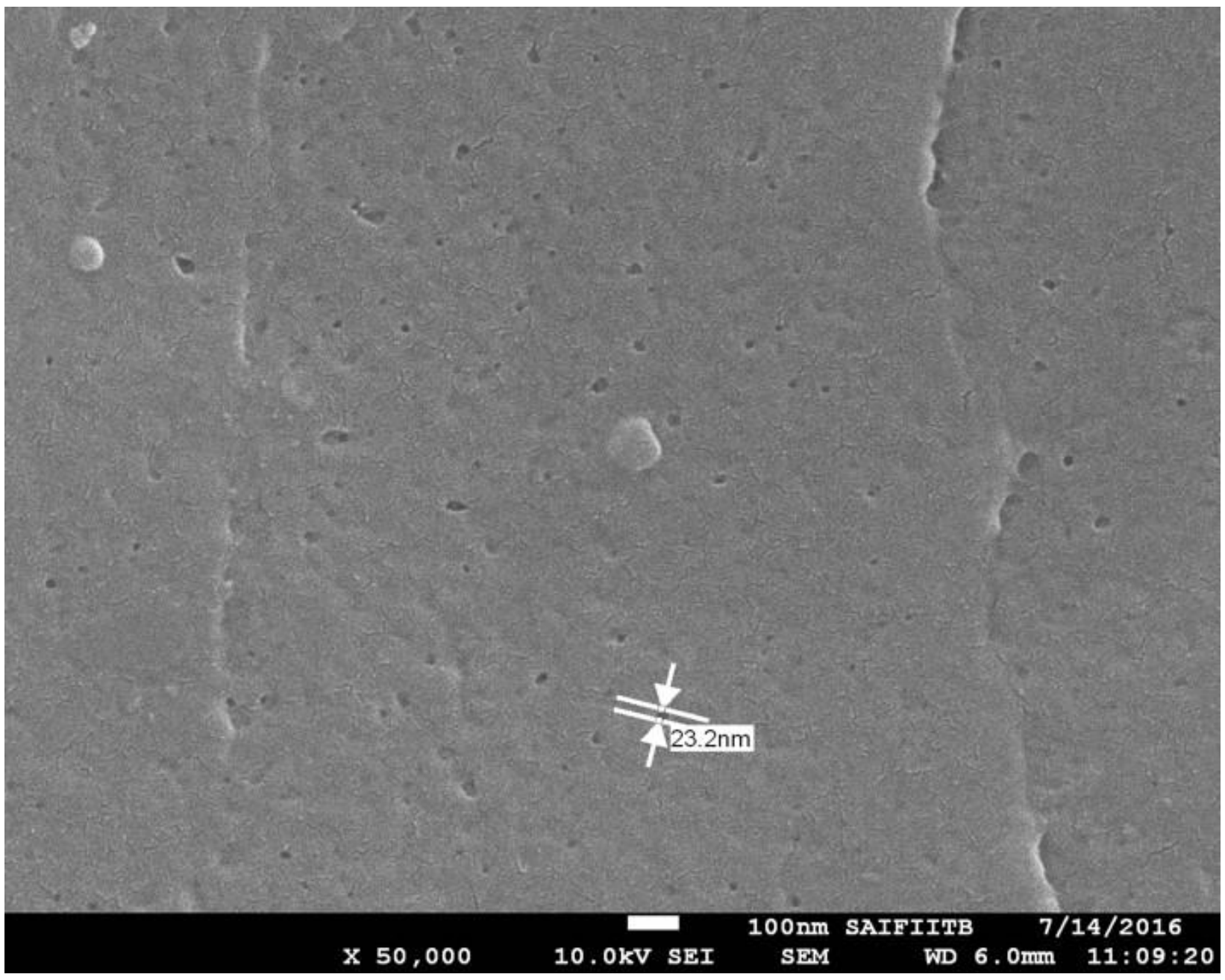

**Fig. 6.  SEM micrograph of pure polysulfone membrane surface**

The pore sizes in above membranes were independently studied using SANS, which is a widely used technique for this purpose[39-41]. Unlike microscopic techniques which provide information about only a small part of sample, SANS gives average information over macroscopic sized sample. The experiments were carried out using SANS Diffractometer at Dhruva reactor[42]. The details of the experiments, data analysis and results of SANS studies are given in earlier paper[37.] The main finding of these studies was that pore sizes are similar in all the membranes. It was seen that average pore size is about 12 nm with polydispersity parameter of 0.4, when data were analyzed in terms of log normal distribution for pore sizes. It is recalled that water permeability of (PSf + MWCNT) membrane is higher than that of pure PSf membrane. The fact that pore sizes in two membranes are similar, it is believed that MWCNTs in (PSf + MWCNT) provide additional path for flow of water. It is not surprising that all membranes have same PEO rejection efficiency, as pore sizes in these membranes are similar. Further SANS studies showed that sizes of $Fe_3O_4$ nanoparticles in the polymer matrix are similar to those in $Fe_3O_4$ powders. This suggests that $Fe_3O_4$ nanoparticles are uniformly dispersed in the membrane and there is no aggregation of these particles.

## 4. Polysulfone-MWCNT nanocomposite membranes with aligned nanotubes

### *4.1 Use of magnetic field for aligning MWCNTs in CNT based nanocomposite membranes*

In general, magnetic field is not suitable field for aligning non-magnetic particles. Sosnick *et al.*[43], however, showed that rod-like nonmagnetic particles can be aligned using moderate magnetic field, if the said particles are dispersed in a suspension of magnetic nano-particles and one applies magnetic field to the suspension. Magnetic forces cause the moments or spins, of the magnetic particles to align along the direction of magnetic field, and their alignment in turn, causes alignment of the rod-like particles. Sosnick *et al*.[43] used above method for aligning rod-like Tobacco Mosaic Virus (TMV) in biological systems. Goyal *et al* [30] pointed out that it should be possible to use above technique for aligning MWCNTs in polymer solutions also. The protocol suggested by Goyal et al. is illustrated in Fig. 7. It involves mixing of MWCNTs and magnetic nano-particles (e.g., oleic acid coated $Fe_3O_4$ particles) to the dope solution (Step-1) and casting of film in presence of magnetic field (Step-2). It is expected that magnetic field would align $Fe_3O_4$ particles along the direction of the magnetic field, and their alignment in turn, would align carbon nano-tubes (Step-3). That is, this method involves casting of membrane in presence of magnetic field.

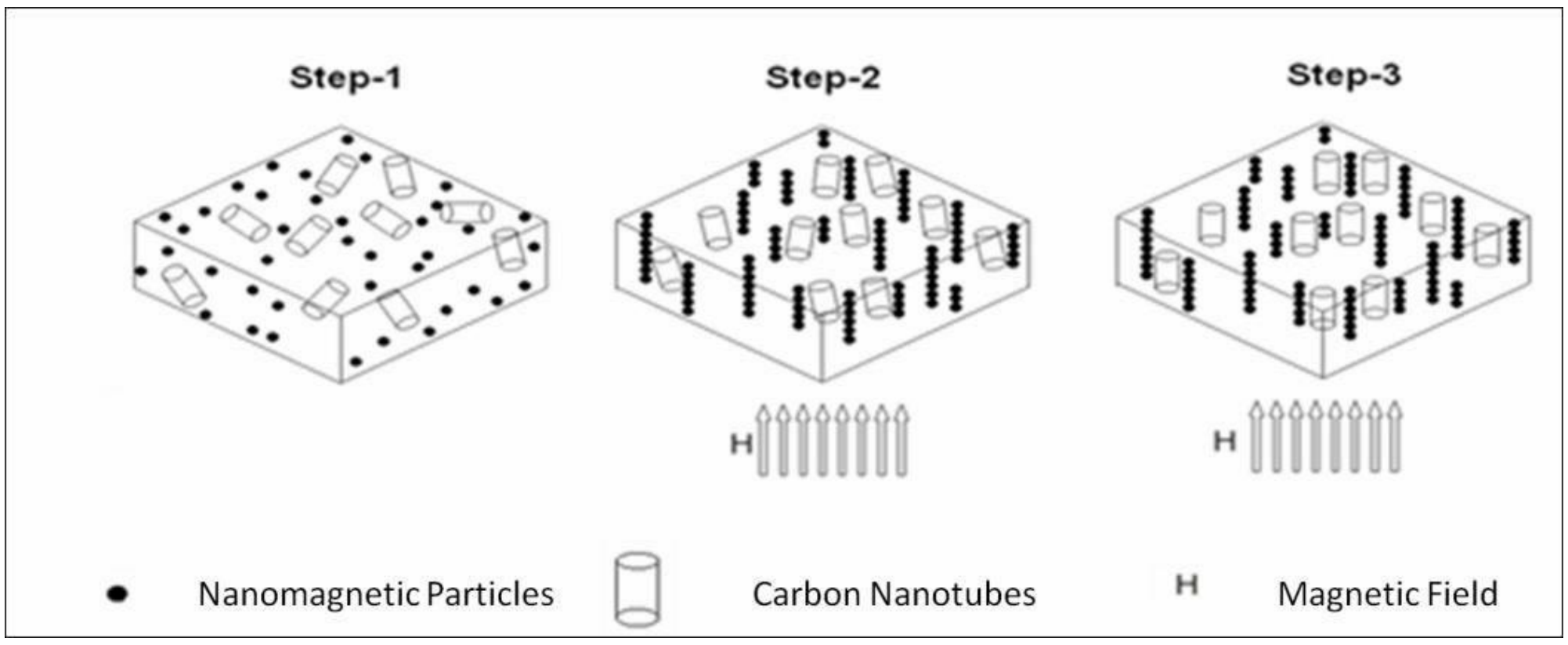

**Fig.7 Schematic representation of steps involved in alignment of CNTs using magnetic nano-particles and magnetic field[30]**

It is important that film is exposed to magnetic field when it is in viscous liquid state. Once the process of phase inversion starts, film would be in a semi-solid form and magnetic field would not be able to change the orientations of MWCNTs. The above method has been used for synthesizing Polysulfone-MWCNT nanocomposite membranes having vertically aligned nanotubes.

### *4.2 Membrane casting machine for casting membranes in magnetic field*

Commercial membrane casting machines, usually, do not have provision to apply magnetic field. We have developed a table top semi-automatic membrane casting machine where it is possible to cast the membrane in presence of magnetic field. Fig.8 is a schematic representation of the machine.

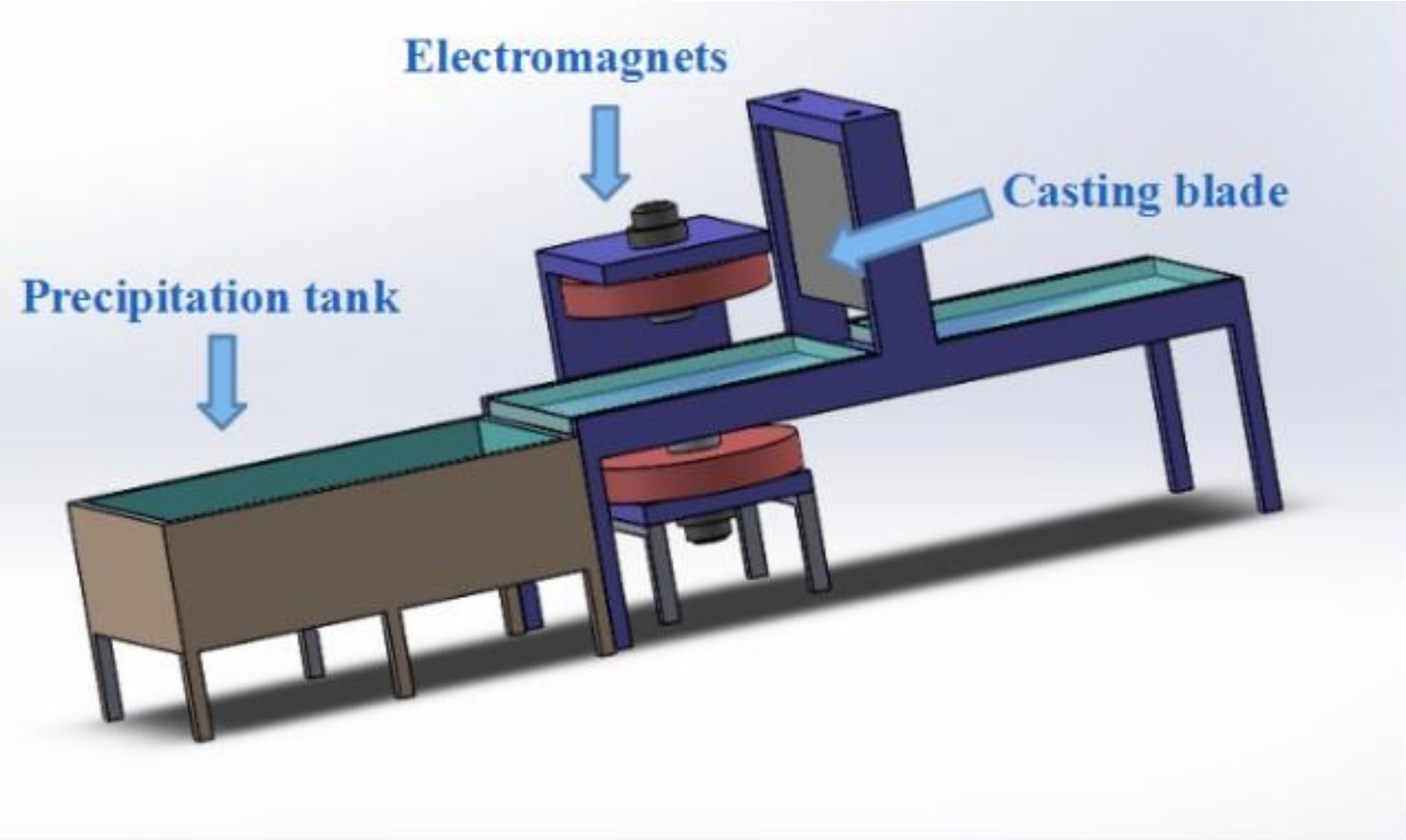

**Fig.8: Schematic representation of membrane casting machine where membrane is casted in presence of magnetic field**

It consists of a casting plate, a casting blade, an electromagnet and a water tank. In the conventional machines, casting plate is fixed and the dope solution is spread on the plate using the casting blade (see Section 3). However, in the present machine, dope solution is spread into a thin film by moving the casting plate. The position of blade is kept fixed as otherwise its movement would be obstructed by the magnet. To ensure smooth movement of the plate, we have used SS casting plate. A special care has been taken to avoid the use of magnetic materials (mild steel etc.) in the fabrication of the machine. Synthesis of membrane in the above design, involves pouring of dope solution on the casting plate, allowing the plate to pass underneath the casting blade and then between the two magnetic poles and finally submerging it in water tank. The time taken in spreading the dope solution in the form of a film is referred to as casting time and the time for which the film is exposed to magnet (or air) before submerging it in water is referred to as evaporation time. It seems, evaporation time and casting time play important role in deciding the quality of casted membrane. For example, it is useful to use larger casting time for more viscous dope solution. Similarly, it will be useful to expose the film to magnetic field for a longer time, when one is aligning MWCNTs using magnetic field. It may be mentioned that one cannot increase evaporation time beyond a limit as, over a time; the film tends to solidify because of its interaction with moisture in the air. In any case, it is desirable that membrane casting machine has a provision to vary both, casting time and evaporation time.

Fig. 9 is the photograph of the machine which have been indigenously fabricated and installed in our laboratory. The major components of the machine are similar to those shown in the schematic drawing (Fig.8). Casting plate and casting blade are made of stainless steel. Casting plate has been machined/ polished to have a smooth surface over an area of about 150 mm x 150 mm within an accuracy of ±10 µm. The magnetic field is applied to the polymer film using an electromagnet, which provides uniform magnetic field over an area of 100 mm x 100 mm. It is possible to vary the magnetic field continuously in range of 0 – 1500 Gauss. Spacing between blade edge and the casting blade can be adjusted to obtain film of required thickness in range of 50 µm to 500 µm. That is, total thickness of the film and substrate, if any, can be up to 500 µm. The operation of the machine has been automated. Once one pours dope solution on the plate and presses the start button, the casting plate slides underneath the blade at a speed which is decided by the preset casting time and then it stays between the pole pieces of magnet for a preset evaporation time. After elapse of evaporation time, the plate slides in to precipitation tank, such that polymer film is fully submerged in the water. That is, one has not to lift and immerse the plate in water tank. As a part of synthesis cycle, film automatically gets immersed in water. This has been achieved by moving the casting plate at an angle to the horizontal surface. The plate can be moved back to the home position with a push of a button and the machine is ready for next cycle. The casting time and evaporation time can be adjusted continuously in range of 1 sec. to 100 sec.

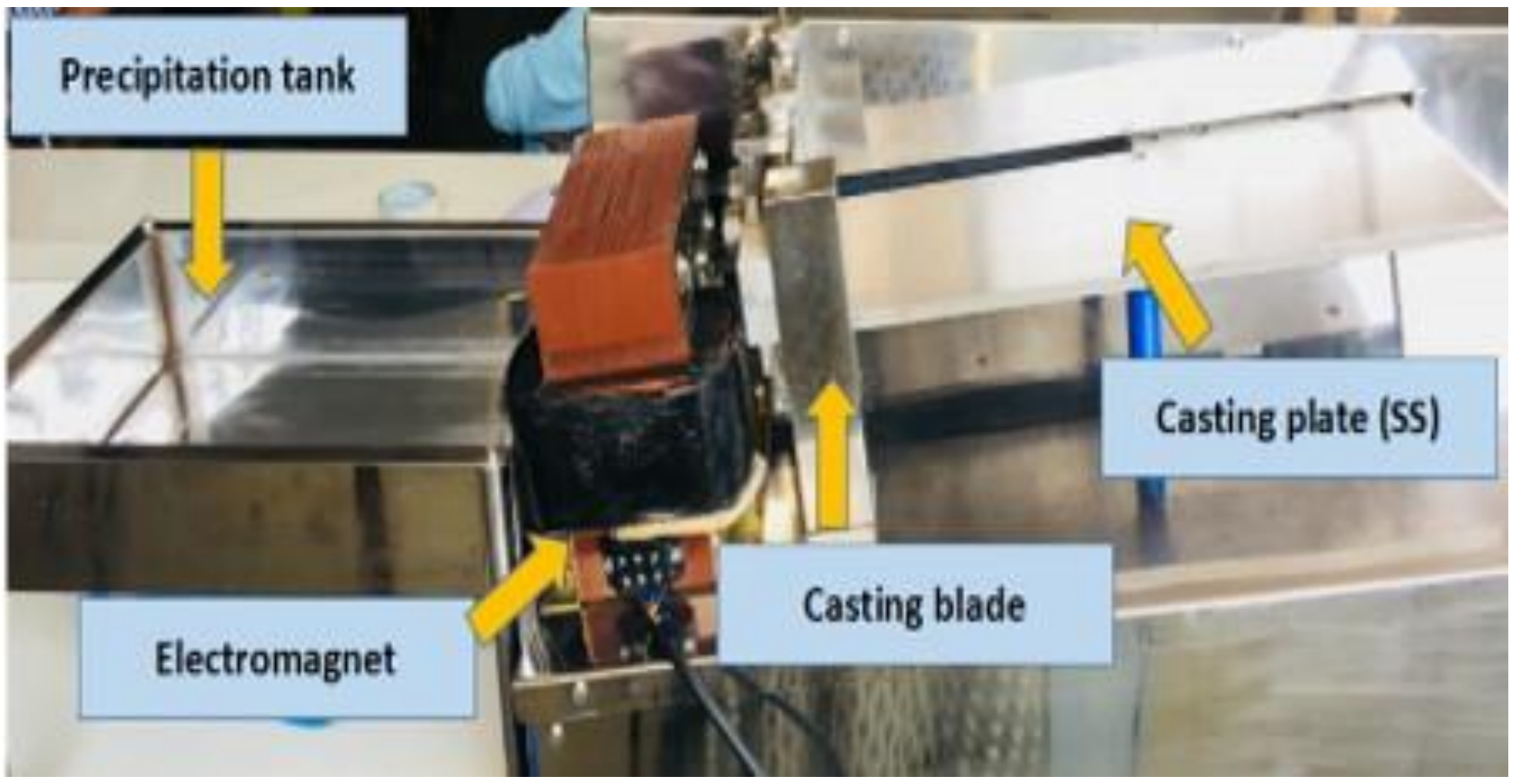

**Fig. 9: Photograph of semi-automatic table top membrane casting machine**

*4.3 Synthesis and water permeability studies on Polysulfone-MWCNT (aligned) membranes.*

Polysulfone-MWCNT nanocomposite membranes containing aligned MWCNTs were synthesized using above membrane casting machine. The alignment of MWCNTs was facilitated by mixing $Fe_3O_4$ nanoparticles in the dope solution and casting the membrane in presence of magnetic field. That is, dope solution consisted of PSf as base polymer, PVP as pore former, NMP as solvent and MWCNTs and $Fe_3O_4$ particles as additives. The relative concentrations of the constituents are given in Table 1. The membrane casting involved pouring of dope solution on the casting plate, allowing the plate to pass underneath the casting blade and then between the two magnetic poles and finally dipping it in water tank. During the time interval when viscous liquid film is exposed to magnetic field, $Fe_3O_4$ particles align along the direction of the magnetic field, and their alignment in turn aligns MWCNTs. This viscous liquid film, having MWCNTs aligned in a direction perpendicular to the membrane surface, solidifies on immersion into water. It is believed that alignment of MWCNTs would not be disturbed during precipitation as the whole process is automated and there is not much time lag between exposing the film to the magnetic field and submerging it in the water. Two sets of membranes were casted. The evaporation time was kept fixed at $\tau$ = 5 for Set-1 membranes (Sr. No. 1 – 5 in Table 2) and magnetic field was increased from 0 to 1500 Gauss. In case of Set-2 (Sr. No. 5 - 7 in Table 2) membranes, the magnetic field as kept fixed at H = 1500 Gauss and the exposure time was varied. All membranes had a fixed thickness of ~100 µm.

Water permeation of above membranes was measured using a cross-flow filtration cell, the details of which are given in Section 3. The results are shown in Table 2. The last column in the Table gives the normalized (= $F / F_0$) water flux corresponding to different experimental conditions. Here $F_0$ is the permeated water flux corresponding H = 0 and $\tau$ = 5 sec, when MWCNTs are not aligned. The normalized flux directly gives information about the factor by which permeability increases because of alignment of MWCNTs.

Results corresponding to Set-1 membranes cleanly show that water permeability of membrane increases from 1.0 to 3.98, when the magnetic field is increased from 0.0 to 1500 Gauss. It is believed increase in water permeation is because of alignment of MWCNTs and the degree of alignment of MWCNTs increases with increase in magnetic field. The degree of alignment or permeated water flux did not saturate at H = 1500 G. The present studies suggest that it should be possible to obtain even higher water permeability if membrane was casted using magnetic fields higher than 1500 Gauss. It is noted that alignment of MWCNTs starts for H > 1000 Gauss.

**Table 2. Effect of Magnetic field on water permeability of the Polysulfone-MWCNT nanocomposite membranes**

| Sr. No. | Composition of membrane (Nominal concentrations of additives) | Evaporation time | Magnetic field (Gauss) | Normalized permeated water flux $F / F_0$ |
|---|---|---|---|---|
| 1 | PSf + CNT (0.1%) + $Fe_3O_4$ (1%) | 5 Sec | 0 | 1.0 |
| 2 | PSf + CNT (0.1%) + $Fe_3O_4$ (1%) | 5 Sec | 500 | 1.0 |
| 3 | PSf + CNT (0.1%) + $Fe_3O_4$ (1%) | 5 Sec | 1000 | 1.0 |
| 4 | PSf + CNT (0.1%) + $Fe_3O_4$ (1%) | 5 Sec | 1200 | 2.35 |
| 5 | PSf + CNT (0.1%) + $Fe_3O_4$ (1%) | 5 Sec | 1500 | 3.98 |
| 6 | PSf + CNT (0.1%) + $Fe_3O_4$ (1%) | 10 Sec | 1500 | 5.65 |
| 7 | PSf + CNT (0.1%) + $Fe_3O_4$ (1%) | 20 Sec | 1500 | 3.91 |

The water permeation studies on Set-2 membranes bring out the role of evaporation time $\tau$ on the properties of casted membranes. It is seen that normalized water flux increases from 3.98 to 5.65 when $\tau$ is increased from 5 sec to 10 sec at a fixed magnetic field of 1500 Gauss. This indicates that the degree of alignment of MWCNTs would be better if the film is exposed to magnetic field for a longer time. However, $\tau$ cannot be increased indefinitely. When the polymer film is exposed to air for a long time, it starts to undergo precipitation because of moisture in the air, and magnetic field would not be able to align the MWCNTs. It is believed that the decrease in F in going from $\tau$ = 10 sec to $\tau$ = 20 sec for H =1500 G is connected with partial precipitation of the film. To verify the above results on alignment of MWCNTs, an effort was made to analyze the structures of above membranes using transmission electron microscope (TEM) and Small Angle X-ray Scattering (SAXS). However, the signal was very poor and it was not possible to find out if MWCNTs are aligned or not. It seems that the contrast between MWCNTs and polymer matrix is very poor, both, for electrons and X-rays.

The present studies show that it is possible to align MWCNTs and achieve higher permeability of polysulfone-MWCNT nanocomposite membranes by casting the membranes in moderate magnetic field. It is clear, however, that the degree of alignment of MWCNTs or water permeability of membrane depends on several parameters, such as concentration of MWCNTs, concentration and sizes of $Fe_3O_4$ particles, viscosity of dope solution, membrane thickness, membrane-casting time, evaporation time, the strength of magnetic field and the evaporation time. No effort has been made to optimize above parameters. All the same, above studies provided a proof of concept for the proposed method of aligning CNTs. (SWCNTs or MWCNTs). This is an important result as alignment of CNTs is important not only for membrane technology but also for several other applications of CNTs. Moreover, present method of aligning CNTs can be easily up scaled for commercial production of CNT base nanocomposite membranes.

## 5. Conclusion

Membrane based processes have got tremendous potential for water treatment applications, that involve low-energy and environmentally friendly operation. This paper deals with development of carbon nanotubes (CNTs) based membranes as it is known that CNT (SWCNTs or MWCNTs) based membranes are more efficient than the conventional membranes. In particular, polysulfone composite membranes having vertically aligned multi-walled carbon nanotubes (MWCNTSs) were synthesized. MWCNTs were aligned by mixing MWCNTs and magnetite ($Fe_3O_4$) nanoparticles to the polymer solution and casting the membranes under the influence of magnetic field. A table top semiautomatic membrane casting machine which allows casting of the membrane in presence of magnetic field, was designed and fabricated. The machine has a provision to vary the magnetic field and the exposure time. The water permeation efficiency of the membranes was measured using a cross-flow filtration cell, which was, again, designed and fabricated indigenously.

There was significant increase in water permeability when membranes were casted in presence of magnetic field. The normalized water flux permeating through the membrane increased from 1.0 to 3.98, when the magnetic field was increased from 0.0 to 1500 Gauss. This increase in water permeability of membrane is connected with the alignment of MWCNTs. Higher is the field, higher would-be degree of alignment of tubes and higher would be the water permeability of the membrane. It was seen that degree of alignment of MWCNTs and water permeability of membrane depend on exposure time also. The normalized water flux permeating through the membrane increased from 3.98 to 5.65 when $\tau$ was increased from 5 sec to 10 sec.

The degree of alignment of MWCNTs could not be determined using TEM or SAXS as there is very poor contrast between MWCNTs and the polymer matrix. It is believed that increase in water permeability and its systematic behavior with change in field or change in exposure time is a good indication of alignment of MWCNTs. In short, this paper reports a new method of aligning CNTs embedded in polymer films and it has potential of up-gradation.

**Acknowledgements**

We gratefully acknowledge the guidance and help of R. C. Bindal, Soumitra Kar and Amit Singha with whom we had interacted in the initial stages of this work. We thank Divya Padmanabhan and Abhishek Narse for their help in designing the electromagnet and Biswajit Panda for his help in synthesis of magnetic nanoparticles. We would also like to thank Priam V. Pillai, Sandeep M. Joshi and R.I.K. Moorthy for useful discussions and encouragement. The financial support from BRNS (Grant no.: 37(2)/14/04/2015-BRNS) is acknowledged.